\pdfoutput=1
\documentclass[12pt,a4paper]{article}

\usepackage{ifthen} 
\usepackage{multirow}
\usepackage{pdflscape}
\usepackage{float}
\usepackage{booktabs}
\usepackage{rotating}
\usepackage{amsmath}
\usepackage{amssymb}
\usepackage{rotating}
\usepackage{verbatim}  
\usepackage[flushleft]{threeparttable}
\usepackage[usenames,dvipsnames]{xcolor}
\definecolor{taplum}{rgb}{0.67843, 0.49804, 0.65882}

\newboolean{pdflatex}
\setboolean{pdflatex}{true} 

\newboolean{articletitles}
\setboolean{articletitles}{true} 

\newboolean{uprightparticles}
\setboolean{uprightparticles}{false} 

\newboolean{inbibliography}
\setboolean{inbibliography}{false} 

\usepackage[top=1in, bottom=1.25in, left=1in, right=1in]{geometry}

\usepackage{microtype}
\usepackage{lineno}  
\usepackage{xspace} 
\usepackage{caption} 

\usepackage{graphicx}  
\usepackage{color}
\usepackage{colortbl}
\graphicspath{{./figs/}} 

\usepackage{amsmath} 
\usepackage{amssymb}
\usepackage{amsfonts}
\usepackage{upgreek} 

\newcommand*\patchAmsMathEnvironmentForLineno[1]{%
\expandafter\let\csname old#1\expandafter\endcsname\csname #1\endcsname
\expandafter\let\csname oldend#1\expandafter\endcsname\csname
end#1\endcsname
 \renewenvironment{#1}%
   {\linenomath\csname old#1\endcsname}%
   {\csname oldend#1\endcsname\endlinenomath}%
}
\newcommand*\patchBothAmsMathEnvironmentsForLineno[1]{%
  \patchAmsMathEnvironmentForLineno{#1}%
  \patchAmsMathEnvironmentForLineno{#1*}%
}
\AtBeginDocument{%
\patchBothAmsMathEnvironmentsForLineno{equation}%
\patchBothAmsMathEnvironmentsForLineno{align}%
\patchBothAmsMathEnvironmentsForLineno{flalign}%
\patchBothAmsMathEnvironmentsForLineno{alignat}%
\patchBothAmsMathEnvironmentsForLineno{gather}%
\patchBothAmsMathEnvironmentsForLineno{multline}%
\patchBothAmsMathEnvironmentsForLineno{eqnarray}%
}

\usepackage{hyperref}    
\usepackage[all]{hypcap} 

\usepackage{xspace} 
\usepackage{upgreek}

\def\lhcb {\mbox{LHCb}\xspace}

\def\babar  {\mbox{BaBar}\xspace}
\def\belle  {\mbox{Belle}\xspace}

\def\MagUp {\mbox{\em Mag\kern -0.05em Up}\xspace}

\ifthenelse{\boolean{uprightparticles}}%
{
 
 \def\Pgamma      {\ensuremath{\upgamma}\xspace}

 \def\Ppi         {\ensuremath{\uppi}\xspace}

 \def\Pphi        {\ensuremath{\upphi}\xspace}

 \def\Ppsi        {\ensuremath{\uppsi}\xspace}

 \def\PDelta      {\ensuremath{\Delta}\xspace}                 
 \def\PXi      {\ensuremath{\Xi}\xspace}                 
 \def\PLambda      {\ensuremath{\Lambda}\xspace}                 
 \def\PSigma      {\ensuremath{\Sigma}\xspace}                 
 \def\POmega      {\ensuremath{\Omega}\xspace}                 
 \def\PUpsilon      {\ensuremath{\Upsilon}\xspace}                 
 
 \def\PB      {\ensuremath{\mathrm{B}}\xspace}                 
                  
 \def\PD      {\ensuremath{\mathrm{D}}\xspace}

 \def\PJ      {\ensuremath{\mathrm{J}}\xspace}                 
 \def\PK      {\ensuremath{\mathrm{K}}\xspace}

 \def\Pb      {\ensuremath{\mathrm{b}}\xspace}

 \def\Pe      {\ensuremath{\mathrm{e}}\xspace}

 \def\Pi      {\ensuremath{\mathrm{i}}\xspace}

 \def\Ps      {\ensuremath{\mathrm{s}}\xspace}

}
{
 
 \def\Pgamma      {\ensuremath{\gamma}\xspace}

 \def\Ppi         {\ensuremath{\pi}\xspace}

 \def\Pphi        {\ensuremath{\phi}\xspace}

 \def\Ppsi        {\ensuremath{\psi}\xspace}                 
                  
 \mathchardef\PDelta="7101
 \mathchardef\PXi="7104
 \mathchardef\PLambda="7103
 \mathchardef\PSigma="7106
 \mathchardef\POmega="710A
 \mathchardef\PUpsilon="7107
                  
 \def\PB      {\ensuremath{B}\xspace}                 
                  
 \def\PD      {\ensuremath{D}\xspace}

 \def\PJ      {\ensuremath{J}\xspace}                 
 \def\PK      {\ensuremath{K}\xspace}

 \def\Pb      {\ensuremath{b}\xspace}

 \def\Pe      {\ensuremath{e}\xspace}

 \def\Pi      {\ensuremath{i}\xspace}

 \def\Ps      {\ensuremath{s}\xspace}

}

\makeatletter
\ifcase \@ptsize \relax
  \newcommand{\miniscule}{\@setfontsize\miniscule{4}{5}}
\or
  \newcommand{\miniscule}{\@setfontsize\miniscule{5}{6}}
\or
  \newcommand{\miniscule}{\@setfontsize\miniscule{5}{6}}
\fi
\makeatother

\DeclareRobustCommand{\optbar}[1]{\shortstack{{\miniscule (\rule[.5ex]{1.25em}{.18mm})}
  \\ [-.7ex] $#1$}}

\def\epem       {{\ensuremath{\Pe^+\Pe^-}}\xspace}

\def\g      {{\ensuremath{\Pgamma}}\xspace}

\def\squark    {{\ensuremath{\Ps}}\xspace}

\def\bquark    {{\ensuremath{\Pb}}\xspace}

\def\pion   {{\ensuremath{\Ppi}}\xspace}
\def\piz    {{\ensuremath{\pion^0}}\xspace}

\def\pip    {{\ensuremath{\pion^+}}\xspace}
\def\pim    {{\ensuremath{\pion^-}}\xspace}

\def\kaon    {{\ensuremath{\PK}}\xspace}
  \def\Kbar    {{\kern 0.2em\overline{\kern -0.2em \PK}{}}\xspace}

\def\KorKbar    {\kern 0.18em\optbar{\kern -0.18em K}{}\xspace}

\def\Kp      {{\ensuremath{\kaon^+}}\xspace}
\def\Km      {{\ensuremath{\kaon^-}}\xspace}

\def\Kstarz  {{\ensuremath{\kaon^{*0}}}\xspace}

  \def\Dbar    {{\kern 0.2em\overline{\kern -0.2em \PD}{}}\xspace}

\def\DorDbar    {\kern 0.18em\optbar{\kern -0.18em D}{}\xspace}

\def\B       {{\ensuremath{\PB}}\xspace}
\def\Bbar    {{\ensuremath{\kern 0.18em\overline{\kern -0.18em \PB}{}}}\xspace}

\def\BorBbar    {\kern 0.18em\optbar{\kern -0.18em B}{}\xspace}
\def\Bz      {{\ensuremath{\B^0}}\xspace}

\def\Bu      {{\ensuremath{\B^+}}\xspace}

\def\Bp      {{\ensuremath{\Bu}}\xspace}

\def\Bd      {{\ensuremath{\B^0}}\xspace}
\def\Bs      {{\ensuremath{\B^0_\squark}}\xspace}
\def\Bsb     {{\ensuremath{\Bbar{}^0_\squark}}\xspace}

\def\jpsi     {{\ensuremath{{\PJ\mskip -3mu/\mskip -2mu\Ppsi\mskip 2mu}}}\xspace}

  \def\Y#1S{\ensuremath{\PUpsilon{(#1S)}}\xspace}

\def\Lbar        {{\ensuremath{\kern 0.1em\overline{\kern -0.1em\PLambda}}}\xspace}
\def\LorLbar    {\kern 0.18em\optbar{\kern -0.18em \PLambda}{}\xspace}

\newcommand{\decay}[2]{\ensuremath{#1\!\to #2}\xspace}         

\def\to                 {\ensuremath{\rightarrow}\xspace}

\newcommand{\tauBz}{{\ensuremath{\tau_{\Bz}}}\xspace}

\def\CP                {{\ensuremath{C\!P}}\xspace}

\newcommand{\dms}{{\ensuremath{\Delta m_{\squark}}}\xspace}

\newcommand{\DGs}{{\ensuremath{\Delta\Gamma_{\squark}}}\xspace}

\newcommand{\Gs}{{\ensuremath{\Gamma_{\squark}}}\xspace}

\newcommand{\ADelta}{{\ensuremath{{\mathcal{A}}^\Delta}}\xspace}

\def\BsToJPsiPhi  {\decay{\Bs}{\jpsi\phi}}
\def\BdToJPsiKst  {\decay{\Bd}{\jpsi\Kstarz}}

\def\BsPhiGam     {\decay{\Bs}{\phi \g}}
\def\BdKstGam     {\decay{\Bd}{\Kstarz \g}}

\def\BdKstee  {\decay{\Bd}{\Kstarz\epem}}

\def\AT#1     {\ensuremath{A_{\mathrm{T}}^{#1}}\xspace}           
\def\btosgam  {\decay{\bquark}{\squark \g}}

\def\C#1      {\ensuremath{\mathcal{C}_{#1}}\xspace}                       
\def\Cp#1     {\ensuremath{\mathcal{C}_{#1}^{'}}\xspace}                    
\def\Ceff#1   {\ensuremath{\mathcal{C}_{#1}^{\mathrm{(eff)}}}\xspace}        
\def\Cpeff#1  {\ensuremath{\mathcal{C}_{#1}^{'\mathrm{(eff)}}}\xspace}       
\def\Ope#1    {\ensuremath{\mathcal{O}_{#1}}\xspace}                       
\def\Opep#1   {\ensuremath{\mathcal{O}_{#1}^{'}}\xspace}                    

\newcommand{\tev}{\ifthenelse{\boolean{inbibliography}}{\ensuremath{~T\kern -0.05em eV}\xspace}{\ensuremath{\mathrm{\,Te\kern -0.1em V}}}\xspace}
\newcommand{\gev}{\ensuremath{\mathrm{\,Ge\kern -0.1em V}}\xspace}
\newcommand{\mev}{\ensuremath{\mathrm{\,Me\kern -0.1em V}}\xspace}
\newcommand{\kev}{\ensuremath{\mathrm{\,ke\kern -0.1em V}}\xspace}
\newcommand{\ev}{\ensuremath{\mathrm{\,e\kern -0.1em V}}\xspace}
\newcommand{\gevc}{\ensuremath{{\mathrm{\,Ge\kern -0.1em V\!/}c}}\xspace}
\newcommand{\mevc}{\ensuremath{{\mathrm{\,Me\kern -0.1em V\!/}c}}\xspace}
\newcommand{\gevcc}{\ensuremath{{\mathrm{\,Ge\kern -0.1em V\!/}c^2}}\xspace}
\newcommand{\gevgevcccc}{\ensuremath{{\mathrm{\,Ge\kern -0.1em V^2\!/}c^4}}\xspace}
\newcommand{\mevcc}{\ensuremath{{\mathrm{\,Me\kern -0.1em V\!/}c^2}}\xspace}

\def\invfb   {\ensuremath{\mbox{\,fb}^{-1}}\xspace}

\def\ps   {\ensuremath{{\mathrm{ \,ps}}}\xspace}

\def\invps{\ensuremath{{\mathrm{ \,ps^{-1}}}}\xspace}

\def\gsim{{~\raise.15em\hbox{$>$}\kern-.85em
          \lower.35em\hbox{$\sim$}~}\xspace}
\def\lsim{{~\raise.15em\hbox{$<$}\kern-.85em
          \lower.35em\hbox{$\sim$}~}\xspace}

\def\pt         {\mbox{$p_{\mathrm{ T}}$}\xspace}

\def\tell1  {TELL1\xspace}
\def\ukl1   {UKL1\xspace}

\usepackage{cite} 
\usepackage{mciteplus}

\usepackage{longtable} 

\begin{document}

\renewcommand{\thefootnote}{\fnsymbol{footnote}}
\setcounter{footnote}{1}


\begin{titlepage}
\pagenumbering{roman}

\vspace*{-1.5cm}
\centerline{\large EUROPEAN ORGANIZATION FOR NUCLEAR RESEARCH (CERN)}
\vspace*{1.5cm}
\noindent
\begin{tabular*}{\linewidth}{lc@{\extracolsep{\fill}}r@{\extracolsep{0pt}}}
\ifthenelse{\boolean{pdflatex}}
{\vspace*{-2.7cm}\mbox{\!\!\!\includegraphics[width=.14\textwidth]{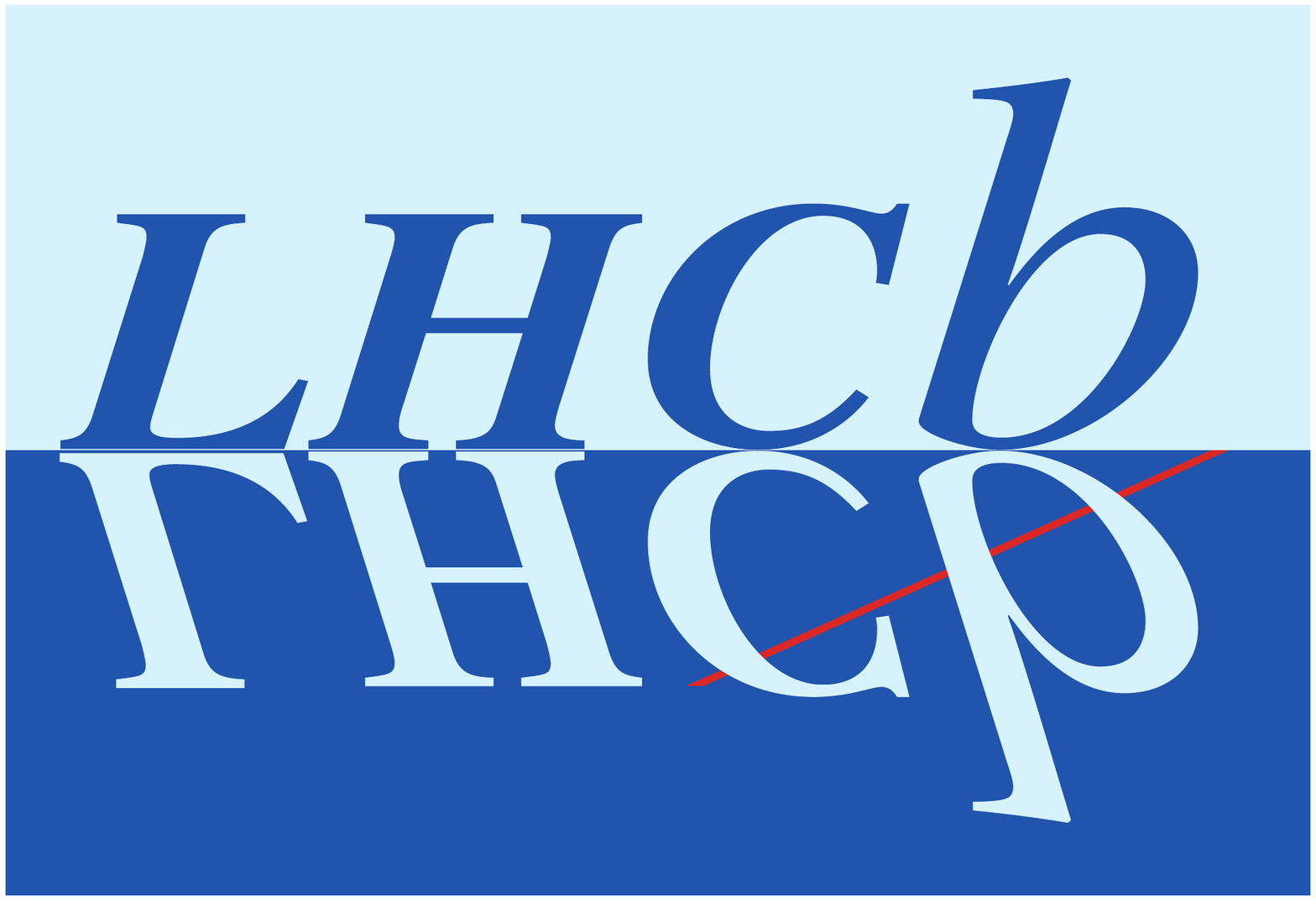}} & &}%
{\vspace*{-1.2cm}\mbox{\!\!\!\includegraphics[width=.12\textwidth]{lhcb-logo.eps}} & &}%
\\
 & & CERN-EP-2016-210 \\  
 & & LHCb-PAPER-2016-034 \\  
 & & February 24, 2017 \\ 
 & &  \\
\end{tabular*}

\vspace*{4.0cm}

{\normalfont\bfseries\boldmath\huge
\begin{center}
  First experimental study of photon polarization in radiative \Bs decays
\end{center}
}

\vspace*{2.0cm}

\begin{center}
The LHCb collaboration\footnote{Authors are listed at the end of this letter.}
\end{center}

\vspace{\fill}

\begin{abstract}

\noindent The polarization of photons produced in radiative \Bs decays 
is studied for the first time.
The data are recorded by the \lhcb experiment in $pp$ collisions 
corresponding to an integrated luminosity of 3\invfb at center-of-mass 
energies of $7$ and $8$\tev.
A time-dependent analysis of the \BsPhiGam decay rate is conducted
to determine the parameter \ADelta, which is related to the ratio 
of right- over left-handed photon polarization 
amplitudes in \btosgam transitions.
A value of
${\mathcal{A}}^\Delta=-0.98^{\,+0.46\,+0.23}_{\,-0.52\,-0.20}$
is measured. 
This result is consistent with the Standard Model prediction within 
two standard deviations. 
 
\end{abstract}

\vspace*{2.0cm}

\begin{center}
  Published in Phys.~Rev.~Lett.~118, 021801 (2017)  
\end{center}

\vspace{\fill}

{\footnotesize 
\centerline{\copyright~CERN on behalf of the \lhcb collaboration, licence \href{http://creativecommons.org/licenses/by/4.0/}{CC-BY-4.0}.}}
\vspace*{2mm}

\end{titlepage}


\newpage
\setcounter{page}{2}
\mbox{~}
\cleardoublepage


\renewcommand{\thefootnote}{\arabic{footnote}}
\setcounter{footnote}{0}



\pagestyle{plain} 
\setcounter{page}{1}
\pagenumbering{arabic}


\newcommand{\ccoef}{\ensuremath{\mathcal{C}}\xspace}
\newcommand{\scoef}{\ensuremath{\mathcal{S}}\xspace}

In the Standard Model (SM), photons emitted in \btosgam transitions 
are produced predominantly with a left-handed polarization, 
with a small right-handed component proportional to the ratio of the quark 
masses, $m_\squark/m_\bquark$.
In many extensions of the SM, the right-handed component can be enhanced, 
leading to observable effects in mixing-induced \CP asymmetries  
and time-dependent decay rates of radiative \Bz and \Bs 
decays~\cite{Atwood:1997zr, Muheim:2008vu}.
Measurements of the time-dependent \CP asymmetries in radiative 
heavy meson decays have been performed by the \babar and \belle 
collaborations in the \Bz system only~\cite{hfag}. 
The production of polarized photons in \btosgam transitions 
was observed for the first time at \lhcb
by studying the up-down asymmetry in \decay{\Bp}{\Kp\pim\pip\g} 
decays~\cite{LHCb-PAPER-2014-001} 
(charge conjugation is implied throughout the text).
In addition, angular observables in the \BdKstee channel 
for dielectron invariant masses of less than $1$\gevcc 
that are sensitive to the polarization of the virtual photon
have also been measured at \lhcb~\cite{LHCb-PAPER-2014-066}. 
All of these measurements are found to be in agreement 
with the SM predictions. 

This Letter reports the first experimental study of the photon polarization 
in radiative \Bs decays, 
determined from the time dependence of the rate of \BsPhiGam decays. 
The rate at which \Bs or \Bsb mesons decay to a common final state 
that contains a photon, such as $\phi \gamma$, depends on
the decay time $t$ and is proportional to
\vspace{-0.7cm}
\begin{center}
\begin{equation}
\label{eq:tagged}
e^{-\Gs t} \big\{ 
          \cosh{\left( \DGs t/2 \right)} 
- \ADelta \sinh{\left( \DGs t/2 \right)}
+ \zeta \, \mathcal{C} \cos{\left( \dms t \right)} 
- \zeta \, \mathcal{S} \sin{\left( \dms t \right)} 
\big\},
\end{equation}
\end{center}
where \DGs and \dms are the width and mass differences 
between the light and heavy \Bs mass eigenstates,
\Gs is the mean decay width,
and $\zeta$ takes the value $+1$ for an initial \Bs state and $-1$ for \Bsb.
The coefficients $\mathcal{C}$, $\mathcal{S}$ and \ADelta 
are functions of the left- and right-handed photon polarization 
amplitudes~\cite{Muheim:2008vu}.
The terms \ccoef and \scoef can be measured only if the initial flavor 
is known: for an approximately equal mixture of \Bs and \Bsb mesons, 
as used in this analysis, these terms cancel and the photon polarization 
affects only the parameter \ADelta.
This approach has the advantage that there is no need to determine 
the flavor of the \Bs candidates at production,
which would considerably reduce the effective size of the data sample.
Compared to the \Bz system, the \Bs is unique in that the sizeable width 
difference allows \ADelta to be measured.
In the SM it can be parameterized as $\ADelta = \sin\left( 2\psi \right)$, 
where 
$\tan \psi \equiv 
 |A(\decay{\Bsb}{\phi \g_\text{R}})|
/|A(\decay{\Bsb}{\phi \g_\text{L}})|$ 
is the ratio of right- and left-handed photon amplitudes. The SM prediction is 
\ensuremath{{\cal A}^\Delta_{\text{SM}}}$ = 0.047 \,^{+ \, 0.029}_{- \, 0.025}$~\cite{Muheim:2008vu}.


This analysis is based on a data sample
corresponding to 3\invfb of integrated luminosity,
collected by the \lhcb experiment
in $pp$ collisions at  center-of-mass energies of 7 and 8\tev
in 2011 and 2012, respectively.
The LHCb detector is a single-arm forward spectrometer 
covering the pseudorapidity range $2 < \eta < 5$, 
described in detail in Refs.~\cite{Alves:2008zz,LHCb-DP-2014-002}.
Different types of charged hadrons are distinguished using 
information from two ring-imaging Cherenkov detectors.
The online event selection is performed by a trigger, 
which consists of a hardware stage, based on information 
from the calorimeter and muon systems, 
followed by a software stage, which applies a full event reconstruction.
Two trigger selections are defined with different photon and track 
momentum thresholds, depending on whether the hardware stage triggered on 
one of the tracks or on the photon. 
Samples of simulated events, produced with the software described in 
Refs.~\cite{Sjostrand:2006za,*Sjostrand:2007gs,LHCb-PROC-2010-056,
Lange:2001uf,Golonka:2005pn,Allison:2006ve,*Agostinelli:2002hh, 
LHCb-PROC-2011-006}, 
are used to characterize signal and background contributions. 

The decay mode \BdKstGam, with \decay{\Kstarz}{\Kp\pim}, is used as 
a control channel.
Since it is a flavor-specific decay, its decay-time distribution is
not sensitive to the photon polarization.
Throughout this Letter, \Kstarz denotes $K^{\ast}(892)^{0}$.  
Candidate \BsPhiGam and \BdKstGam decays are reconstructed from a photon, 
and two oppositely charged tracks: 
two kaons to reconstruct \decay{\phi}{\Kp\Km} decays and
a kaon and a pion to reconstruct \decay{\Kstarz}{\Kp\pim} decays. 
The selection is designed to maximize the expected significance 
of the signal yield. 
Photons are reconstructed from energy deposits in the electromagnetic 
calorimeter and are required to have momentum transverse to the beam axis, 
\pt, larger than 3.0\gevc or 4.2\gevc, depending on the trigger selection. 
Each charged particle is required to have a minimum \pt of 0.5\gevc 
and at least one of them must have \pt larger than 1.7\gevc or 1.2\gevc, 
depending on the trigger selection. 
The tracks are required to be inconsistent with originating 
from a primary $pp$ interaction vertex.  
The pion and kaon candidates are required to be identified 
by the particle identification system.
The two tracks must meet at a common vertex and have 
an invariant mass within 15\mevcc of the known $\phi$ 
mass~\cite{Agashe:2014kda} for the signal mode, 
or within 100\mevcc of the known \Kstarz mass for the control mode. 
Each \Bs or \Bd candidate is required to have \pt larger than 3.0\gevc, 
and a reconstructed momentum consistent with originating from 
one and only one primary vertex.
Background due to photons from \piz decays 
is rejected by a dedicated algorithm~\cite{CalvoGomez:2042173}.
In addition, the cosine of the helicity angle, 
defined as the angle between the positively charged hadron 
and the \B meson in the rest frame of the $\phi$ or \Kstarz meson, 
is required to be less than 0.8. 

A kinematic fit of the full decay chain is performed, 
imposing a constraint on the mass of the \B candidate. 
Its decay time is determined from the fitted four-momentum 
and flight distance from the primary vertex. 
The mass constraint improves the decay-time resolution 
and also ensures that it is not correlated with 
the reconstructed mass for the signal. 
Only candidates with decay times between 0.3 and 10 ps are retained.


\begin{figure}[tb]
    \centering
    \includegraphics[width=8.6cm]{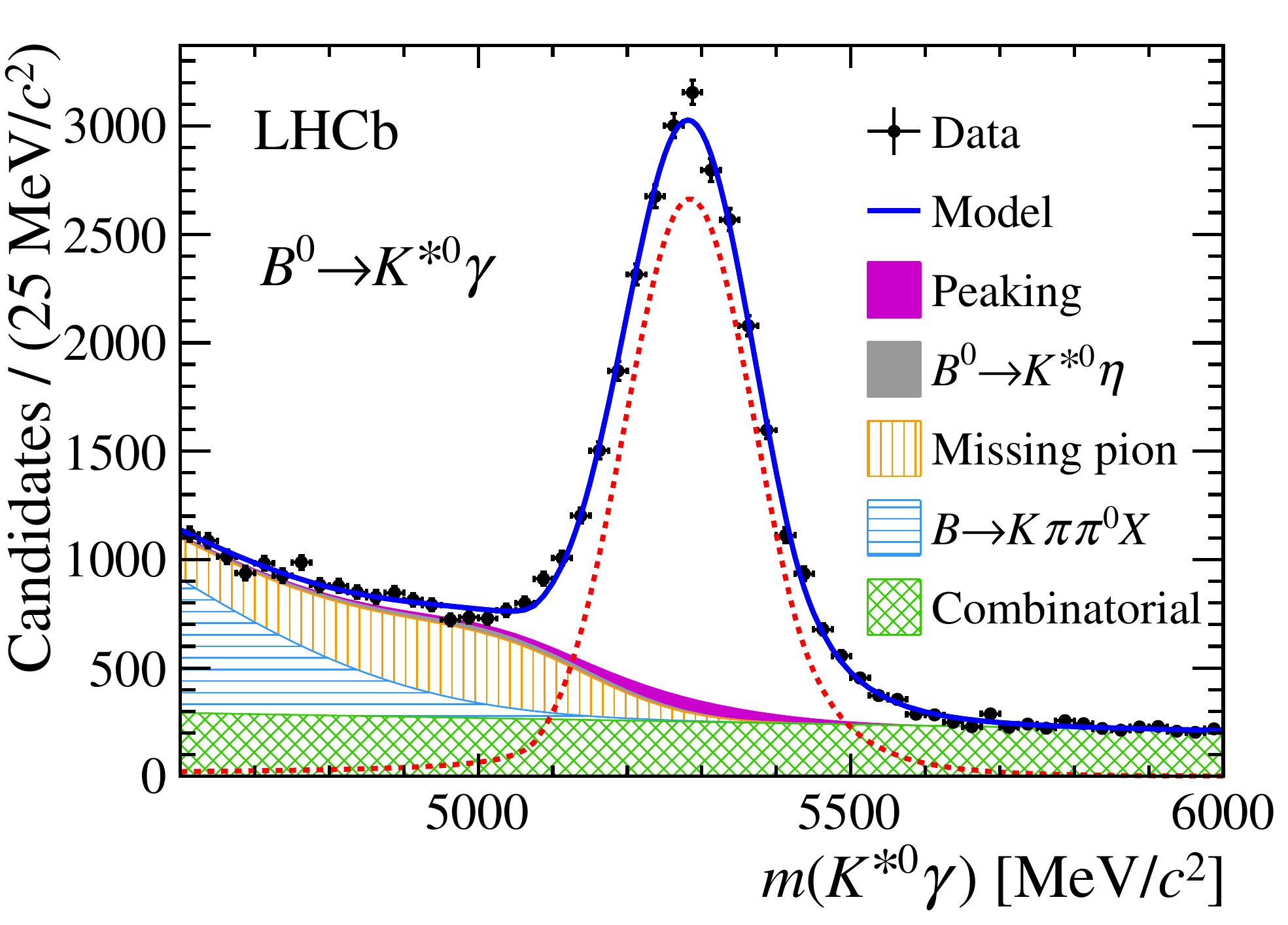} 
    \includegraphics[width=8.6cm]{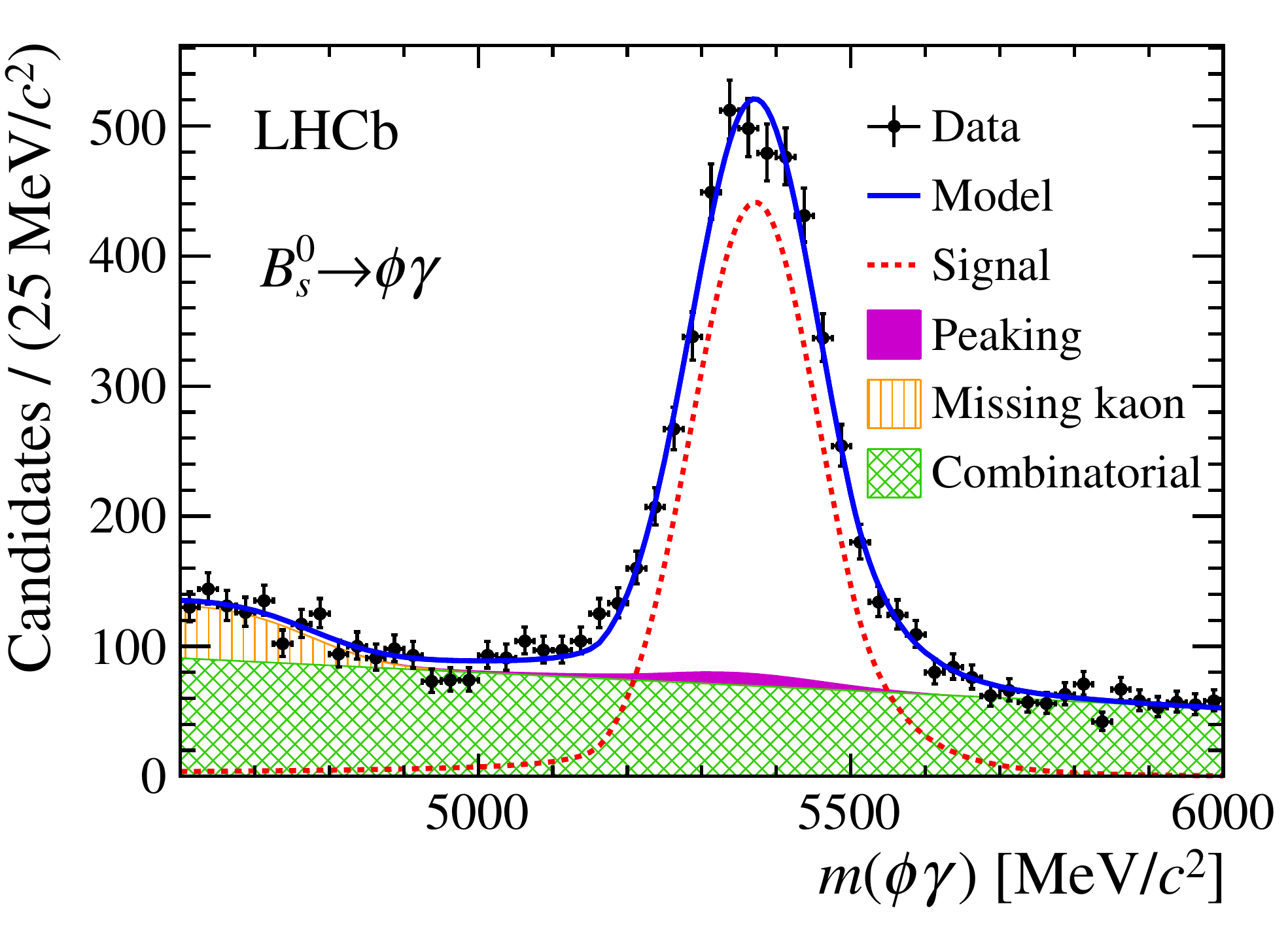}
    \caption{Fits to the invariant mass distributions
of the \Bz (top) and \Bs (bottom) candidates.}
    \label{fig:mass}
\end{figure}

The \Bs and \Bz signal yields are obtained from separate 
extended unbinned maximum likelihood fits to the \Pphi\g and \Kstarz\g 
invariant mass distributions, shown in Fig.~\ref{fig:mass}.
The signal line shapes are described by modified 
Crystal Ball functions~\cite{Skwarnicki:1986xj} 
with tails on both sides of the peak.
The tail parameters are determined from simulation.
Three background categories are considered: 
peaking, partially reconstructed and combinatorial backgrounds.
Peaking backgrounds are due to the misidentification of a final-state particle. 
All possible sources of misidentified tracks, 
as well as misidentification of a \piz meson as a photon, 
are considered for the signal and control channels.   
Partially reconstructed backgrounds, 
in which one or more final-state particles are not reconstructed, 
are described with an ARGUS function~\cite{argus} 
convolved with a Gaussian function 
to account for the mass resolution of the detector.
The dominant contributions are decays with a missing pion or kaon, 
$\B \to K \pi \piz X$, and $\Bz \to \Kstarz \eta$. 
All shape parameters for the peaking and partially reconstructed backgrounds 
are fixed from simulation.
The ratios of the yields of peaking backgrounds to signal are fixed 
using previous measurements~\cite{Agashe:2014kda,LHCb-PAPER-2012-019}.
A first-order polynomial is used to describe the combinatorial background. 
The signal yields are $4072\pm112$ and $24\,808\pm321$ 
for the \decay{\Bs}{\Pphi\g} and \decay{\Bz}{\Kstarz\g} decays, 
where the uncertainties are statistical only.

main    

The mass fits are used to assign each candidate
of the \BsPhiGam and \BdKstGam samples a signal weight
to subtract the backgrounds~\cite{Pivk:2004ty}.
An unbinned maximum likelihood fit
of the weighted decay-time distributions~\cite{Xie:2009rka}
is then performed simultaneously on the \BsPhiGam and \BdKstGam samples. 
The signal probability density function (PDF) is defined 
from the product of the decay-time-dependent signal rate, $\mathcal{P}(t)$,
and the efficiency, $\epsilon(t)$,
convolved with the resolution.

\newpage

For \BsPhiGam, Eq.~\ref{eq:tagged} reduces to 
\vspace{-0.7cm}
\begin{center}
\begin{equation}
\mathcal{P}(t) \propto e^{-\Gs t} 
\big\{ 
          \cosh{\left(\DGs t/2 \right)} 
- \ADelta \sinh{\left(\DGs t/2 \right)} 
\big\} 
\end{equation}
\end{center}
when summing over initial \Bs and \Bsb states.
The \Bs and \Bsb production rates are assumed to be equal,
given that their measured asymmetries~\cite{LHCb-PAPER-2014-042} 
are found to have a negligible effect on the measurement of \ADelta. 
For \BdKstGam, the decay-time-dependent signal rate is a single exponential 
function, $\mathcal{P}(t) \propto e^{-t_{}^{} / \tauBz}$.
The physics parameters \tauBz, \Gs, and \DGs
are constrained to the averages from Ref.~\cite{hfag}: 
$\tauBz = 1.520 \pm 0.004$\ps, 
$\Gs    = 0.6643 \pm 0.0020$\invps, and
$\DGs   = 0.083 \pm 0.006$\invps. 
The correlation of $-0.239$ between the uncertainties 
on \Gs and \DGs is taken into account.

To ensure that the simulation reproduces the decay-time resolution, 
additional control samples of \BsToJPsiPhi and \BdToJPsiKst decays are used,
where the \jpsi meson is reconstructed from a pair of oppositely charged muons. 
Selections mimicking those of \BsPhiGam and \BdKstGam, 
treating the \jpsi meson as a photon, are applied. 
The distributions of the difference in position 
between the reconstructed \jpsi and \Pphi or \Kstarz vertices 
are measured in data and simulation and found to be in agreement.
The decay-time-dependent resolution functions are then 
determined from the simulation. 
The decay-time resolution is small compared to the \bquark-hadron lifetimes,
and similar for \BsPhiGam and \BdKstGam.

\begin{figure}[tb]
    \centering
    \includegraphics[width=8.6cm]{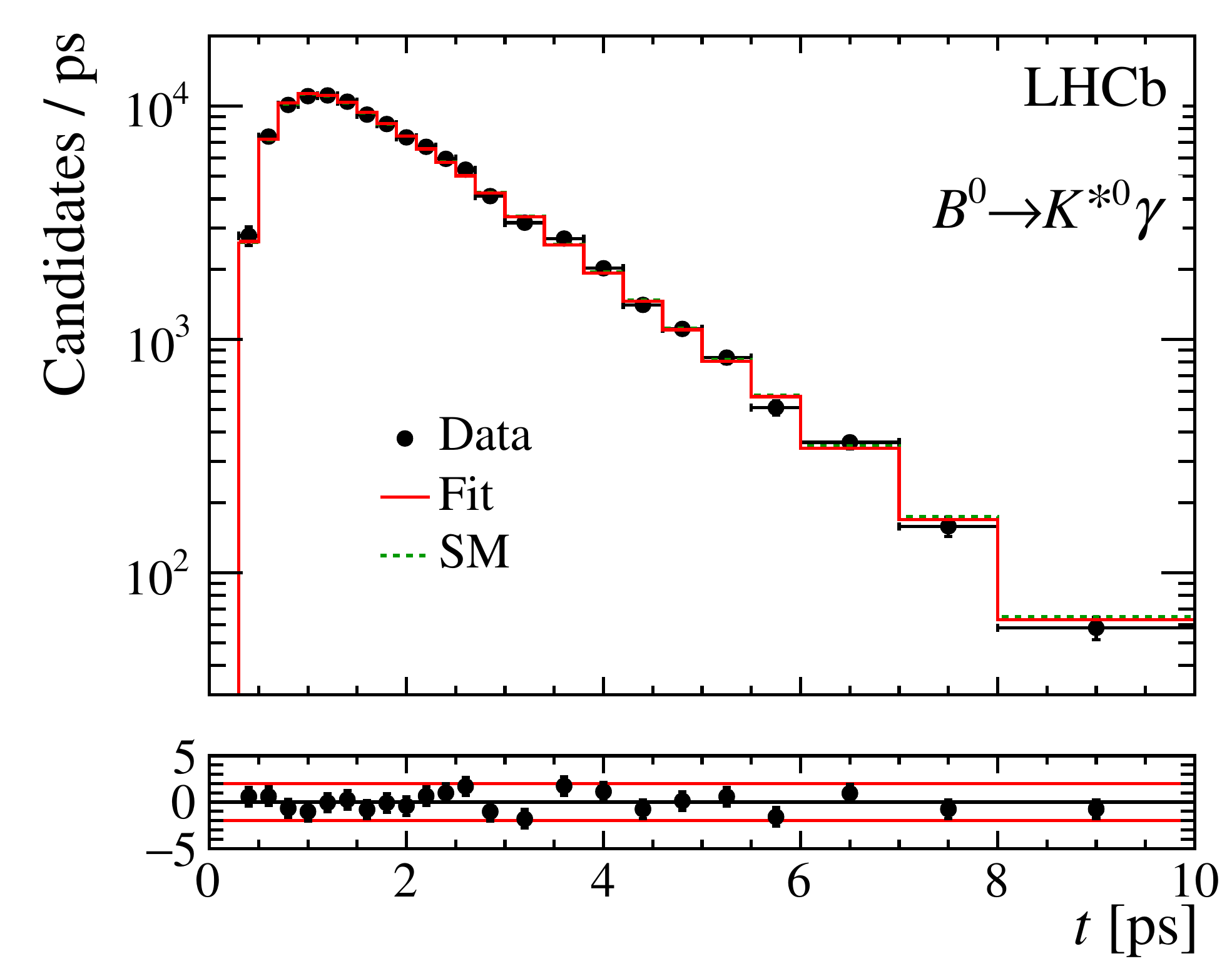} 
    \includegraphics[width=8.6cm]{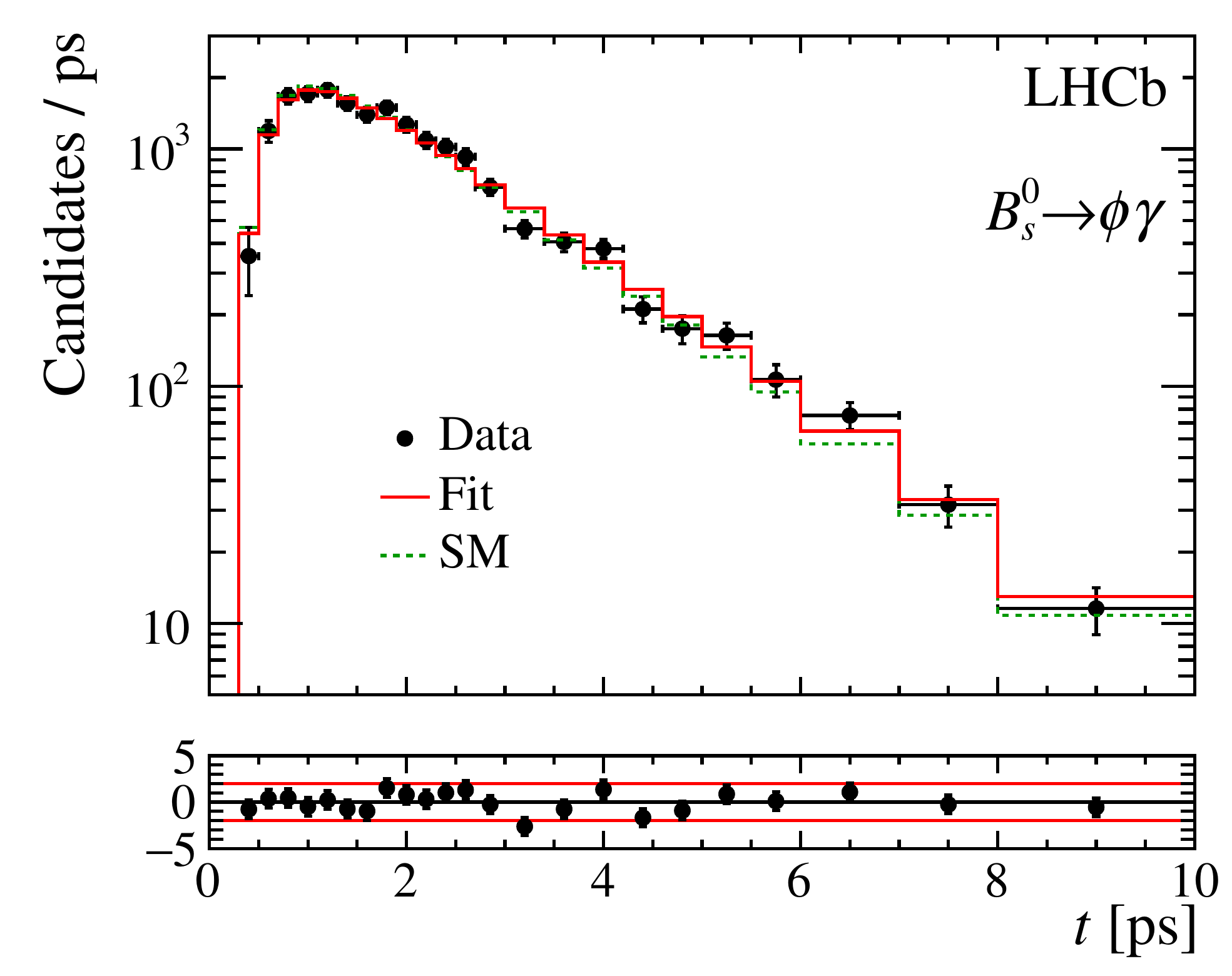}
    \caption{Background-subtracted decay-time distributions
for \BdKstGam (top) and \BsPhiGam (bottom) decays 
with the fit projections overlaid 
and normalized residuals shown below. 
The projections of a fit with \ADelta fixed to the central value of 
the SM prediction~\cite{Muheim:2008vu} are also shown.
For display purposes, the PDF is shown as a histogram, 
integrated across each decay-time interval.}
    \label{fig:directfit}
\end{figure}

\newpage

The decay-time-dependent efficiency is parameterized as
\vspace{-0.7cm}
\begin{center}
\begin{equation}
\epsilon(t) = e^{-\alpha t} \,  \frac{[a \, (t-t_0)]^n}{1 + [a \, (t-t_0)]^n} 
\ \ {\rm for} \ t \ge t_0,
\end{equation}
\end{center}
where the parameters $a$ and $n$ describe the curvature 
of the efficiency function at low decay times, 
$t_0$ is the decay time below which the efficiency function is zero,
and $\alpha$ describes the decrease of the efficiency at high decay times.
Large simulated samples of \BsPhiGam or \BdKstGam decays 
are used to validate this parameterization. 
The signal PDF is found to describe the reconstructed decay-time distribution
of selected simulated candidates over the full decay-time range. 
The \BsPhiGam and  \BdKstGam  decay-time-dependent efficiency parameters 
are found to be similar.
In a simultaneous fit of both simulation samples, requiring the parameters 
$a$ and $n$ to be the same for both channels does not change 
the quality of the fit.
To assess whether the simulation reproduces the decay-time-dependent 
efficiency, the \BdKstGam data sample alone is used to fit \tauBz,
fixing in this case all the efficiency parameters to those from the simulation. 
The fitted value of \tauBz is $1.524 \pm 0.013$\ps, 
where the uncertainty is statistical only, 
in agreement with the world average value \cite{hfag}.
In the simultaneous fit to the data, the parameters $a$ and $n$ are 
required to be the same for both channels 
and fixed to their values in the simulation.
For $t_0$ and $\alpha$, a global offset, the same for both channels, 
is allowed between data and the simulation.

Pseudoexperiments are used to validate the overall fit procedure. 
For each pseudoexperiment, samples of \BsPhiGam and \BdKstGam
candidates are generated, including both signal and background contributions.
The expected yields are taken from the fit to the data,
as is the signal mass shape.
Background events are generated according to
the mass and decay-time PDFs determined from fits 
to samples of events generated with the full \lhcb simulation.
For each pseudoexperiment, the mass fits to the \BsPhiGam and \BdKstGam
samples are performed, followed by the decay-time fit to 
the background-subtracted samples. 
The procedure is tested in samples of pseudoexperiments
generated with different values of \ADelta.
No bias on the average fitted value of \ADelta is observed.
Statistical uncertainties are found to be underestimated
by an amount that depends on \ADelta; the effect is  
5.8\% for the value seen in data
and is accounted for in the results below.

The \BdKstGam and \BsPhiGam 
background-subtracted decay-time distributions 
and the corresponding fit projections are shown in 
Fig.~\ref{fig:directfit}. 
The fitted value of \ADelta is $ -0.98 \,^{ + \, 0.46}_{ - \, 0.52}$. 
The statistical uncertainty includes a contribution due to 
the uncertainties on the physics parameters \tauBz, \Gs and \DGs, 
which is estimated to account for $^{+ \, 0.10}_{- \, 0.17}$.

\begin{figure}[tb]
    \centering
    \includegraphics[width=8.6cm]{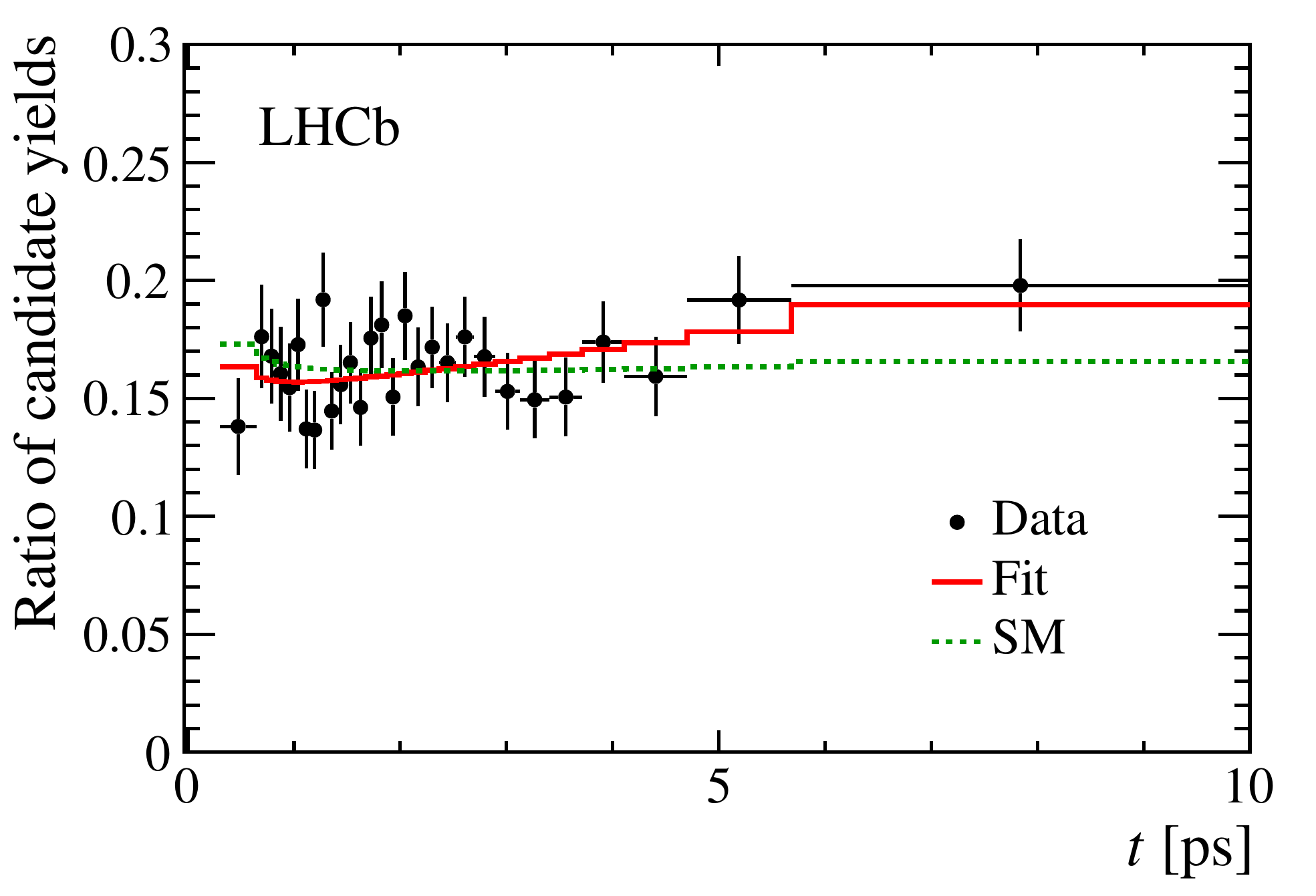} 
    \caption{Decay-time dependence of the ratio of the yields 
of \BsPhiGam and \BdKstGam, with the fit overlaid.
The expected distribution for the central value of the SM 
prediction~\cite{Muheim:2008vu} is also shown.}
    \label{fig:ratiofit}
\end{figure}

In an alternative approach, 
\ADelta is calculated from the 
ratio of the yields of \BsPhiGam and \BdKstGam
in bins of decay time.
Based on a study of pseudoexperiments, 
the binning scheme is designed to have the same number
of events in each bin, thereby optimizing the overall sensitivity to \ADelta.
Decay-time-dependent efficiency and resolution effects are taken into account 
by calculating correction factors in each bin before fitting for \ADelta. 
Pseudoexperiments are used to validate this approach 
and to test its sensitivity, 
which is found to be equivalent to that of the baseline procedure.
The fit to the data is shown in Fig.~\ref{fig:ratiofit}, 
along with the expected distribution for the central value
of the SM prediction for \ADelta.
The fitted value is $\ADelta = -0.85 \,^{+ \, 0.43}_{- \, 0.46}$.
The statistical uncertainty is strongly correlated with that
of the baseline approach;
the difference between the two results is well within the range
expected from pseudoexperiments.


The dominant systematic uncertainty comes from the background subtraction.
It is evaluated to be $^{+ \, 0.19}_{- \, 0.20}$ and includes contributions
from potential correlations between the reconstructed mass 
and decay time for the backgrounds ($\pm 0.15$),
uncertainties on the peaking background yields $(^{+ \, 0.02}_{- \, 0.05})$,
and the models used in the mass fit.
The latter is assessed by the use of alternative models: 
an asymmetric Apollonios function~\cite{Santos:2013gra} 
for the signal ($\pm 0.03$), 
an exponential for the combinatorial background $(\pm 0.07)$,
and several shape variations for the most relevant partially 
reconstructed backgrounds ($\pm 0.10$).
The systematic uncertainty due to the limited size of
the simulation samples used to assess the decay-time-dependent efficiency 
is $^{+ \, 0.13}_{- \, 0.05}$.  
The uncertainties related to the decay-time resolution are negligible. 
The sum in quadrature of these systematic uncertainties is 
$^{+ \, 0.23}_{- \, 0.20}$.


In summary, the polarization parameter \ADelta is measured in 
the first time-dependent analysis of a radiative \Bs decay, 
using a data sample corresponding to 
an integrated luminosity of 3\invfb 
collected by the \lhcb experiment.
This parameter is related to the ratio of right- over left-handed 
photon polarization amplitudes in \btosgam transitions.
More than 4000 \BsPhiGam decays are reconstructed. 
The decay-time-dependent efficiency is calibrated 
with a control sample of \BdKstGam decays that is six times larger.
From an unbinned simultaneous fit to the \BsPhiGam and \BdKstGam data samples, 
a value of
\begin{center}
$\ADelta = -0.98 \,^{+ \, 0.46}_{- \, 0.52} \,^{+ \, 0.23}_{- \, 0.20}$
\end{center}
is measured,
where the first uncertainty is statistical and the second systematic. 
The result is compatible with the SM expectation,
\ensuremath{{\cal A}^\Delta_{\text{SM}}}$= 0.047 \,^{+ \, 0.029}_{- \, 0.025}$~\cite{Muheim:2008vu},
within two standard deviations.

\section*{Acknowledgements}
 
\noindent We express our gratitude to our colleagues in the CERN
accelerator departments for the excellent performance of the LHC. We
thank the technical and administrative staff at the LHCb
institutes. We acknowledge support from CERN and from the national
agencies: CAPES, CNPq, FAPERJ and FINEP (Brazil); NSFC (China);
CNRS/IN2P3 (France); BMBF, DFG and MPG (Germany); INFN (Italy); 
FOM and NWO (The Netherlands); MNiSW and NCN (Poland); MEN/IFA (Romania); 
MinES and FASO (Russia); MinECo (Spain); SNSF and SER (Switzerland); 
NASU (Ukraine); STFC (United Kingdom); NSF (USA).
We acknowledge the computing resources that are provided by CERN, IN2P3 (France), KIT and DESY (Germany), INFN (Italy), SURF (The Netherlands), PIC (Spain), GridPP (United Kingdom), RRCKI and Yandex LLC (Russia), CSCS (Switzerland), IFIN-HH (Romania), CBPF (Brazil), PL-GRID (Poland) and OSC (USA). We are indebted to the communities behind the multiple open 
source software packages on which we depend.
Individual groups or members have received support from AvH Foundation (Germany),
EPLANET, Marie Sk\l{}odowska-Curie Actions and ERC (European Union), 
Conseil G\'{e}n\'{e}ral de Haute-Savoie, Labex ENIGMASS and OCEVU, 
R\'{e}gion Auvergne (France), RFBR and Yandex LLC (Russia), GVA, XuntaGal and GENCAT (Spain), Herchel Smith Fund, The Royal Society, Royal Commission for the Exhibition of 1851 and the Leverhulme Trust (United Kingdom).

\addcontentsline{toc}{section}{References}
\setboolean{inbibliography}{true}
\bibliographystyle{LHCb}
\bibliography{main,LHCb-PAPER,LHCb-CONF,LHCb-DP,LHCb-TDR}

\newpage                                                                                                        

\newpage

\centerline{\large\bf LHCb collaboration}
\begin{flushleft}
\small
R.~Aaij$^{40}$,
B.~Adeva$^{39}$,
M.~Adinolfi$^{48}$,
Z.~Ajaltouni$^{5}$,
S.~Akar$^{6}$,
J.~Albrecht$^{10}$,
F.~Alessio$^{40}$,
M.~Alexander$^{53}$,
S.~Ali$^{43}$,
G.~Alkhazov$^{31}$
% ...

{\footnotesize \it
% ...
$ ^{5}$Clermont Universit{\'e}, Universit{\'e} Blaise Pascal, CNRS/IN2P3, LPC, Clermont-Ferrand, France\\
$ ^{6}$CPPM, Aix-Marseille Universit{\'e}, CNRS/IN2P3, Marseille, France\\
% ...
$ ^{10}$Fakult{\"a}t Physik, Technische Universit{\"a}t Dortmund, Dortmund, Germany\\
% ...
$ ^{31}$Petersburg Nuclear Physics Institute (PNPI), Gatchina, Russia\\
% ...
$ ^{39}$Universidad de Santiago de Compostela, Santiago de Compostela, Spain\\
$ ^{40}$European Organization for Nuclear Research (CERN), Geneva, Switzerland\\
% ...
$ ^{43}$Nikhef National Institute for Subatomic Physics, Amsterdam, The Netherlands\\
% ...
$ ^{48}$H.H. Wills Physics Laboratory, University of Bristol, Bristol, United Kingdom\\
% ...
$ ^{53}$School of Physics and Astronomy, University of Glasgow, Glasgow, United Kingdom\\
% ...
}
\end{flushleft}

\end{document}